\documentstyle[11pt,newpasp,twoside]{article}
\def\edcomment#1{\iffalse\marginpar{\raggedright\sl#1\/}\else\relax\fi}
\reversemarginpar

\begin{document}
\title{Polytropes: Implications for Molecular Clouds and
Dark Matter}
 \author{Christopher F. McKee}
\affil{Departments of Physics and of Astronomy, University of
California, Berkeley CA 94720}

\begin{abstract}
Molecular clouds are supported against their own self-gravity by
several different sources of pressure:  thermal pressure, mean magnetic
pressure, and turbulent pressure.  
Multi-pressure polytropes, in which each of these
pressures is proportional to a power of the density,
can account for many of the observed properties
of molecular clouds.
The agreement with observation can be improved with
composite polytropes, in which an isothermal
core is embedded in a turbulent envelope.
Observed molecular clouds generally have ${\gamma_p}<1$,
corresponding to a velocity dispersion that increases with scale.
For such clouds the ratio of the mean pressure to the surface
pressure must be less than 4.

Small, very dense ($\bar n_{\rm H}\sim 10^{11}$
cm$^{-3}$) molecular clouds have been proposed as models for both dark
matter and for extreme scattering events.  
Insofar as the equation of state in these clouds
can be represented by a single polytropic relation,
such models conflict with observation.  It is
possible to contrive composite polytropes that do
not conflict with observation, but whether the thermal
properties of the clouds are consistent with such structures remains
to be determined.

\end{abstract}


\section{Introduction}

	Understanding star formation
requires understanding 
the interstellar molecular clouds out of which stars form.
These molecular clouds are objects of fascinating dynamical
complexity, exhibiting highly supersonic motions while at
the same time being gravitationally bound (Zuckerman \& Evans
1974; Larson 1981).  The nonthermal motions help support
the clouds against the force of gravity;
thermal pressure and magnetic
fields also contribute.  As is characteristic
of turbulent motions, the amplitude of
the nonthermal motions increases with scale
(Larson 1981).  Since
the nonthermal motions are largest on the 
scale of the cloud itself, they lead to
substantial changes in the shape of the
cloud over time (Ballesteros-Paredes, Vazquez-Semadeni, \&
Scalo 1999).  Attempting to model such a
complex system is a daunting task that
is only now beginning to be tackled numerically
(see Vazquez-Semadeni et al 2000).

     In order to treat the structure of
molecular clouds analytically, it is
necessary to make a number of
approximations.  First, we assume
that the cloud is in a steady state.
The steady-state approximation is plausible since the lifetime
of molecular clouds is typically about
an order of magnitude greater than
their free-fall time (Blitz \& Shu 1980).
For example, Williams \& McKee (1997) have shown that
massive stars will destroy a cloud of mass $M\sim 10^6~M_\odot$ 
by photoionization in $3\times 10^7$ yr; smaller clouds
live longer.  By comparison,
the typical cloud of that mass has a mean density
$\bar n_{\rm H}\simeq 84M_6^{-1/2}$ cm$^{-3}$ (Solomon et al 1987),
corresponding to a free-fall time
$t_{\rm ff}=1.37\times 10^6 (\bar n_{\rm H}/10^3$~cm$^{-3})^{-1/2}$~yr
$=4.7\times 10^6 (M/10^6 \ M_\odot)^{1/4}$ yr.
Clouds of mass $M\la 10^6~M_\odot$ thus live
$\ga 6t_{\rm ff}$.  The steady-state assumption 
is quite approximate, however, since the process
of destroying the cloud by photoionization is very violent.

	Since molecular clouds are approximately in
a steady state and they are gravitationally bound,
it follows that, when averaged over time, they
are approximately in hydrostatic equilibrium.
For example, the majority of
cores in high latitude cirrus clouds 
observed by Turner (1993) are consistent with 
being in hydrostatic equilibrium.

      	Next, we assume that the time-averaged cloud is
spherical.  Some effects that
could lead to non-spherical clouds, including
tidal gravitational fields (Scoville \& Sanders
1987) and rotation (Goodman et al 1993), are observed
to be relatively weak.
Some molecular clouds are observed to be highly filamentary
(e.g., Alves et al 1998), and it has been suggested
that this can be explained by helical magnetic
fields (Fiege \& Pudritz 2000).  Many molecular clouds
are not highly filamentary, however;
for example, only about 15\% of the clouds in
the catalog of Solomon et al (1987),
which is based on $^{13}$CO observations, have aspect ratios exceeding 2.

\section{Polytropes}

	Molecular clouds in the Galaxy 
are observed to be gravitationally bound, 
and we are approxmating them as being spherical and in
hydrostatic equilibrium.  Just as in the case of stars,
it is convenient to model them as polytropes, which
satisfy the equation of hydrostatic equilibrium
\begin{equation}
\frac{dP}{dr} =-\frac{GM\rho}{r^2} ,
\end{equation}
with
\begin{equation}
P(r)=K_p\rho^{{{\gamma_p}}}.
\end{equation}

	The {\it structure} of a polytrope is determined by the polytropic
index
\begin{equation}
{\gamma_p}\equiv 1+\frac{1}{n} ,
\end{equation}
where $n$ is the index used to characterize polytropes in the
theory of stellar
structure.  The {\it stability} of a polytrope is determined
by ${\gamma_p}$ and the adiabatic index $\gamma$ that describes
how the pressure of a mass element responds to a density perturbation,
\begin{equation}
\delta\ln P\equiv \gamma\delta\ln\rho.
\end{equation}

	Polytropes provided the first quantitative model for stars.
In order to have zero pressure at the boundary, which is assumed
to be at a finite radius, it is necessary to have ${\gamma_p}>6/5$.
Stars are stable against gravitational collapse for $\gamma>4/3$.
Since the dynamical timescale for a star is short
compared to the heat flow time, the entropy is constant
in each mass element during a dynamical perturbation;
such perturbations are described as {\it locally adiabatic}
by McKee \& Holliman (1999; hereafter MH).

	Lynden-Bell \& Wood (1968) modeled globular clusters
as bounded, isothermal gas spheres (${\gamma_p}=1$).  Since there
are no internal degrees of freedom, the adiabatic index
is $\gamma=5/3$; as a result, the ``gas'' is non-isentropic.
Globular clusters are {\it globally adiabatic} (MH) since
the heat flow time is comparable to the dynamical time
of the cluster.  As Lynden-Bell \& Wood (1968) showed,
a pressure-bounded cloud with ${\gamma_p}=1,\ \gamma=5/3$
is subject to core collapse when the density
contrast between the center and the surface is too large,
$\rho_c/\rho_s>389.6$ (MH).

\section{Polytropic Models of Molecular Clouds}

	Molecular clouds are supported by three pressure components,
thermal, magnetic, and turbulent.  Lizano \& Shu (1989)
developed the first model that accounted for these three
pressure components.
They assumed that the gas is
isothermal and that the turbulent pressure 
scales as the logarithm of the density (a ``logatrope'');
the magnetic field was assumed to be axisymmetric.  

	MH developed the theory of {\it multi-pressure polytropes} 
in which each pressure component has 
arbitrary values of ${\gamma_p}$ and $\gamma$.  
Clouds with $\gamma<{\gamma_p}$ are convectively unstable,
and were not considered. Holliman (1995) showed that
the effects of an axisymmetric magnetic field 
in which the flux is frozen to the gas could
be approximated by a gas with $\gamma=4/3$.  For a
flux-to-mass distribution corresponding to a uniform
field in a spherical cloud, ${\gamma_p}=4/3$; ambipolar diffusion
reduces ${\gamma_p}$ below 4/3.  For the turbulent pressure,
MH focused on the case of Alfv\'enic turbulence,
which (at least when it is weak) can be modeled with 
${\gamma_p}=1/2,\ \gamma=3/2$ (McKee \& Zweibel 1995).
Since ${\gamma_p}<1$ for Alfv\'en waves, the velocity
dispersion $\sigma\propto \rho^{({\gamma_p}-1)/2}$ increases
with scale, as observed (Larson 1981). 
However, since the Alfv\'en waves are globally
adiabatic, they are less effective at providing
stability than would be expected for 
$\gamma=3/2$; in fact, they have the same stability properties
as a locally adiabatic
component with $\gamma\la 1$. 
The equation of state for molecular clouds
is therefore {\it soft}:  These clouds usually
have ${\gamma_p}<6/5$ unless the magnetic field has
${\gamma_p}> 6/5$ and is sufficiently strong.  Furthermore, all the
pressure components have $\gamma\leq 4/3$
(in the case of Alfv\'en waves, this is
the equivalent locally adiabatic index).
As a result, {\it stable} molecular clouds must be
confined by the pressure of the ambient medium,
and their properties are determined by conditions
at the surface.  For example, stable clouds with
${\gamma_p}<4/3$ satisfy
\begin{equation}
M\leq 4.555\frac{\sigma_s^4}{(G^3P_s)^{1/2}}
\end{equation}
for any value of $\gamma$,
where $\sigma_s$ is the 1D velocity dispersion
at the cloud surface.
For ${\gamma_p}=1$, the coefficient 4.555 is reduced
to 1.182, the value for the Bonnor-Ebert sphere.
The mean pressure in molecular clouds is limited by
the surface pressure: For ${\gamma_p}\leq 1$, it is less than
$4P_s$.  However, the central pressure can
become arbitrarily large compared to the
surface pressure if $\gamma$ is sufficiently greater
than ${\gamma_p}$.  Holliman (1995) showed that multi-pressure
polytropes could successfully account for a number
of the observed properties of molecular clouds.

	The molecular cloud cores in which low-mass stars form
often exhibit central regions that are supported
primarily by the pressure of an isothermal gas, with 
nonthermal motions becoming important only in the envelopes.
To model this two component structure it is convenient
to introduce {\it composite polytropes} (Curry \& McKee 2000)
of the type used for stars many decades ago (e.g., Milne 1930).
Curry \& McKee (2000) showed that composite polytropes
are very promising as models for low-mass cores:
In particular, it is possible to have an isothermal
core with a non-isentropic ($\gamma>{\gamma_p}$) polytropic
envelope in which the central temperature
and the surface pressure are fixed (as they generally
are in practice), but in which the mass is arbitrarily
large.  Such models are consistent with observations
of small NH$_3$ cores in large $^{13}$CO envelopes.
Curry (in preparation) has made detailed comparisons
of composite polytrope models with the observations.

\section{Molecular Clouds as Dark Matter}

	A number of authors have suggested that self-gravitating
clouds of cold molecular gas could account for the dark
matter in the Galactic halo (Pfenniger, Combes, \& Martinet 1994;
Pfenniger \& Combes 1994;
De Paolis et al 1996, 1998; Gerhard \& Silk 1996; Combes \& Pfenniger 1997).
Henriksen \& Widrow (1995) and Walker \& Wardle (1998; hereafter
WW98) went on
to suggest that such clouds would have ionized surfaces and
could therefore account for the ``extreme scattering events''
(ESEs) observed by Fiedler et al (1994).\footnote{The intensity
of the interstellar ionizing radiation field is quite weak
(Reynolds 1984; Slavin, McKee, \& Hollenbach 2000),
however, and to date there is no physically self-consistent
calculation that demonstrates that the surfaces of
these clouds can be sufficiently ionized to account for the
ESEs.}
Henriksen \& Widrow (1995) pointed out that such clouds could cause
gravitational microlensing if the clouds had masses of order 0.1
$M_\odot$.  Draine (1998) showed that gas clouds can also act as
optical lenses and thereby mimic microlensing.

	The clouds considered in the papers by Pfenniger, Combes and
Martinet are assumed to be turbulent, with a hierarchical internal
structure that terminates on the smallest scale in ``clumpuscules''
of mass $\sim 10^{-3}\, M_\odot$ and radius $\sim 30$ AU.
If the clumpuscules exist near the edge of the cloud, so that they are 
exposed to typical interstellar conditions, then they are subject to
the difficulties described below.  In any case, the simulations reviewed
in Vazquez-Semadeni et al (2000) show that turbulent, isothermal
clouds dissipate their internal kinetic energy in
about a dynamical time ($\sim 10^3$ yr), far less than the
$\ga 1$ Gyr cloud lifetimes assumed by these authors.

	Gerhard \& Silk (1996) have shown that clouds embedded in
non-baryonic halos can be in stable hydrostatic equilibrium
even in the absence of a confining medium.  Here we shall argue that
it is unlikely that highly pressured, self-gravitating clouds without
such halos can exist in the Galactic halo.  Valentijn \& van der Werf
(1999) claim to have detected enough molecular hydrogen
in the nearby spiral NGC 891 to account for the dark
matter within the optical disk of that galaxy; the densities they infer
($n_{\rm H}\la 10^3$ cm$^{-3}$) are orders of magnitude less than
those in the models discussed above, however, and do
not violate the arguments presented here.

	The argument is strongest if the clouds can be represented
as single component polytropes.  First consider the case in which
${\gamma_p}\leq 1$.  This includes the case of isothermal clouds
(${\gamma_p}=1$), such as those considered by DePaolis et al (1998) and
by Wardle \& Walker (1999).\footnote{Wardle (private communication)
has pointed out that he and Walker did not assume that the clouds
are isothermal, but rather that they could be characterized
by a mean temperature.  However, since the temperatures
they considered are so close to that of the microwave background
radiation, which sets a floor on the temperature of each part
of the cloud, their models are in fact quite close to being isothermal.}
Such clouds must have a mean pressure
less than 4 times the ambient pressure (MH).  Now, the ambient 
pressure in the halo is less than that in the disk of the Galaxy,
which is $P/k\simeq 2\times 10^4$ K cm$^{-3}$ (Boulares \& Cox 1990;
I have omitted the pressure due to cosmic rays, since they penetrate
into the cloud).  For a given value of the mean cloud density
$\bar n_{\rm H}$, a lower limit on the mean pressure of the cloud is given by
assuming that the cloud is composed of molecular hydrogen
(plus a cosmic abundance of helium) and has a temperature equal to
that of the cosmic microwave background, 2.73~K: 
$\bar P/k > 0.6\times 2.73\bar n_{\rm H}$.  The constraint that this
be less than $4\times 2\times 10^4$ K cm$^{-3}$ gives an upper limit on
the mean cloud density, $\bar n_{\rm H}<5\times 10^4$ cm$^{-3}$.  
(This constraint is comfortably satisfied by galactic molecular 
clouds---see \S 1.)  If the clouds
are polytropes with ${\gamma_p}<1$, this constraint rules out the model
of WW98, who assumed $\bar n_{\rm H}\sim 10^{12}$ cm$^{-3}$.  It also rules out
much of the parameter space considered by DePaolis et al (1998), 
$10^4$ cm$^{-3}$ $\la\bar n_{\rm H}\la 10^8$ cm$^{-3}$.

	Next, consider polytropes with $1<{\gamma_p}< 6/5$.  For such clouds,
$\bar n_{\rm H}/n_s$ increases toward infinity as ${\gamma_p}\rightarrow 6.5$, but the
ratio remains modest unless ${\gamma_p}$ is very close to 6/5.
For example, for ${\gamma_p}=1.15$ one finds $\bar n_{\rm H}/n_s<13.6$.  Hence,
unless ${\gamma_p}$ is extremely close to 6/5, such polytropes cannot
describe the dense clouds proposed to account for the dark
matter.

	Finally, consider polytropes with ${\gamma_p}\geq 6/5$.
Such polytropes can have zero pressure at the boundary,
like stars do, so the above constraint does not apply.
Nonetheless, there are two significant difficulties.
First, since $T\propto \rho^{{\gamma_p}-1}$, the 
large ratio of the mean density to the surface density in these clouds
implies a correspondingly large temperature ratio.  The
maximum density at the surface is for a molecular gas at
2.73 K, as described above; this gives $n_s\leq 1.2\times 10^4$
cm$^{-3}$.  Draine (1998) has calculated polytropic models for clouds
of mean density $\bar n_{\rm H}=6\times 10^{10}$ cm$^{-3}$ with a range of
polytropic indexes from ${\gamma_p}=11/9\simeq 1.222$ (just above the critical
value of 6/5) to ${\gamma_p}=5/3$.  For ${\gamma_p}= 11/9$, the cloud is very
centrally concentrated:  It has a central density $n_c=6190\bar n_{\rm H}
=3.7\times 10^{14}$ cm$^{-3}$,  corresponding to a central
to surface temperature ratio of $T_c/T_s\geq 214$.
In the opposite case of ${\gamma_p}=5/3$, the cloud
is much less centrally concentrated ($n_c/\bar n_{\rm H}=6.0$),
but the temperature ratio is much larger because
of the greater value of ${\gamma_p}$, $T_c/T_s\geq 0.97\times 10^5$.
Setting $T_s=2.73$ K gives $T_c\geq 580,\ 2.6\times 10^5$ K
for ${\gamma_p}=11/9,\ 6/5$, respectively.
(These values are larger than Draine's since he did not impose
the requirement that $T_s\geq 2.73 K$; to make his models
consistent with this constraint, one would
either have to drop the assumption that the cloud
is a single-component polytrope [see below]
or increase $M/R\propto T$ significantly.)
Over the entire range of polytropic indexes
considered by Draine (1998), it is difficult, if not
impossible, to maintain the high central
temperatures required.

	The dominant heating mechanisms that have
been suggested for these clouds are cosmic
rays and X-rays (DePaolis et al 1998 have also
suggested heating by embedded binaries, but we
shall not consider that mechanism here).  In both
cases, the heating decreases inward, making 
it difficult to have a polytropic structure
in which the temperature increases inward,
as it does for ${\gamma_p}>1$.
More significantly, the column density
to the center of a polytrope with ${\gamma_p}\geq 6/5$
is so large that the
heating rate is substantially reduced there.
Cosmic rays are more penetrating that X-rays,
so I shall focus on them.
For the ${\gamma_p}=5/3$ model, I find
$\Sigma_c\equiv \int_0^R \rho dr
=59$ g cm$^{-2}$.
A lower limit on the interaction 
cross section of the cosmic rays is the
nuclear cross section; for relativistic
cosmic rays this is about 0.02 cm$^{2}$ g$^{-1}$
(Webber, Lee, \& Gupta 1992). Thus, in
this case (which has already been 
eliminated by its extremely high central
temperature), the cosmic ray
intensity is reduced by at least a factor
$e$.  The case with the lowest 
central temperature (${\gamma_p}=11/9$) is
much more centrally concentrated,
and as a result it has such a large
central column density ($\Sigma_c=5000$ g cm$^{-2}$) 
that it is impossible for
cosmic rays to reach the center.
We conclude that it is impossible for
dense, but non-stellar, gas clouds that
obey a polytropic equation of state to exist
in the diffuse ISM, much less the
Galactic halo.

	This conclusion can be circumvented 
by dropping the assumption that the cloud
is a single-component polytrope.  A composite
polytrope in which 
the core (with most of the mass) has ${\gamma_p}>6/5$ but the
envelope (with most of the volume) is isothermal can be developed that
is consistent with Draine's model and has
$T_s>2.73$ K (Draine, private communication).  
The central column density of these models
is similar to that of the single component models,
however, so one is restricted to cases with ${\gamma_p}$
close to 5/3 if cosmic rays are to be able to
heat the central regions of the cloud.
Whether such a contrived model is physically realizable
remains to be seen.

	We see that the condition of hydrostatic equilibrium
places severe constraints on the viability of models
of dense gas clouds that have been advanced to account
for dark matter.  Simple polytropic models can be
ruled out, but detailed calculations of the structure
of such clouds with accurate heating and cooling are
necessary to determine if they can exist at the low
pressures characteristic of the diffuse ISM.

\acknowledgments{I wish to thank Charles Curry, Bruce Draine,
David Hollenbach, Chris
Matzner, and Mark Wardle for helpful comments.  
My research is supported
in part by NSF grant AST95-30480 and by a grant from NASA
to the Center for Star Formation Studies.

\end{document}